\documentclass[letterpaper,twocolumn,10pt]{article}

\usepackage[utf8]{inputenc}
\usepackage[T1]{fontenc}
\usepackage{mathptmx}

\usepackage{amsmath,amssymb}
\usepackage{array}
\usepackage{booktabs}
\usepackage{graphicx}
\usepackage{tabularx}
\usepackage[para,online,flushleft]{threeparttable}
\usepackage{subcaption}
\usepackage{authblk}

\usepackage{dirtytalk}
\usepackage{url}
\usepackage{xcolor}
\usepackage{tikz}
\usetikzlibrary{arrows.meta,shapes.geometric}

\usepackage[hidelinks]{hyperref}

\newcommand{\orcid}[1]{%
  \href{https://orcid.org/#1}{\textsuperscript{\textsc{orcid}}}%
}

\tikzset{
    sem-variable/.style={
        draw,
        ellipse,
        align=center,
        minimum width=3cm,
        minimum height=1.5cm,
        font=\small
    },
    sem-path/.style={
        ->,
        >=Stealth,
        line width=0.6pt
    },
    sem-coefficient/.style={
        fill=white,
        inner sep=1.5pt,
        font=\small
    }
}

\newcommand{\authorssource}{Source: Created by the authors.}
\newcommand{\missingfigure}[1]{%
    \fbox{\parbox[c][4cm][c]{0.9\linewidth}{\centering
        Figure file not available:\\[0.5ex]\texttt{#1}}}%
}

\title{Chat First, Worry Later: Understanding Individuals' Privacy Perceptions Using ChatGPT in a Work Context}

\author{%
Christoph Nirschl$^{1}$\,
\quad
Magdalena Glas$^{1}$\,%
\thanks{Corresponding author:
\href{mailto:magdalena.glas@ur.de}{\texttt{magdalena.glas@ur.de}}}
\quad
Gerhard Messmann$^{2}$\,
\quad
Günther Pernul$^{1}$\,
\\
\small $^{1}$University of Regensburg, Faculty of Informatics and
Data Science, Universitätsstr.\ 31, 93040 Regensburg, Germany
\\
\small $^{2}$University of Regensburg, Faculty of Human Sciences,
Universitätsstr.\ 31, 93040 Regensburg, Germany
}

\begin{document}

\date{}
\maketitle

\section*{Abstract}
Generative Artificial Intelligence (GenAI) tools like ChatGPT, which can generate human-like responses from vast amounts of textual data, are increasingly transforming work routines across various fields, including education, healthcare, and IT. This integration, however, raises privacy concerns and questions the readiness of both environments and individuals. To investigate this issue, we conducted a user study with $N=224$ participants from a range of different employment sectors that have integrated ChatGPT into their work routines. We examined how proficiency in the utilization of ChatGPT, general privacy concerns, and organizational policies for GenAI usage impact users' actual ChatGPT usage and how these factors interact. Our findings reveal organizational policies are significantly positively associated with privacy-related ChatGPT proficiency, however, the overall proficiency is low. Higher privacy concerns were found to negatively influence both the frequency of ChatGPT use and the diversity of its applications, especially among users in organizations without GenAI policies.

\section{Introduction}
\label{sec:introduction}
Over the past decade, the field of Artificial Intelligence (AI) has seen remarkable advancements~\cite{Lu19}, which have been pivotal in addressing complex challenges such as data analysis, pattern recognition, and decision-making~\cite{Gupta23}. This progress has been significantly accelerated by the evolution and ongoing development of Large Language Models (LLMs)~\cite{Teubner23}, marking a new era in AI technology known as GenAI~\cite{Goodfellow20}. LLMs are deep neural networks trained on vast corpora of text, enabling them to capture the underlying structures of natural and structured language~\cite{Sandoval23}. By recognizing patterns within this data, LLMs can automatically generate coherent and contextually relevant content based on the knowledge they've acquired~\cite{Guo23}. The substantial impact of this advancement is evident in the emergence and widespread adoption of tools such as OpenAI's \textit{ChatGPT}, Google's \textit{Gemini}, and DeepSeek, LLM-based chatbots capable of generating detailed, human-like responses based on prompt instructions~\cite{Zhao23}. In this study, we focused on a single GenAI tool to ensure comparability in measuring users' proficiency regarding how input data is stored and processed, as these aspects vary across different tools. We selected ChatGPT because it is by far the most widely used GenAI tool, with more than 500 million weekly users as of spring 2025 \cite{Paris2025ChatGPT1B}.
Individuals not only use ChatGPT for private tasks, but also to help them accomplish work-related tasks more efficiently~\cite{Kobiella2024_ChatGPTYoungProfessionals}. Prior research has explored the potential of integrating ChatGPT in various work environments such as education~\cite{Park2024_ChatGPTEducation,Han2024_ChatGPTElementary}, healthcare~\cite{Calle2024_AIDrivenHealthcare}, and IT~\cite{Kabir2024_ChatGPTStackOverflow}, demonstrating their utility in streamlining tasks and facilitating workflows. However, the process of building, training, and refining the underlying model, heavily relies on large amounts of data, predominantly obtained from online sources, including user-generated content~\cite{Peris23}. Given the potential inclusion of sensitive or personal information within this data, concerns regarding privacy and data sensitivity are raised~\cite{Zhao23}. Prior research has shown that ChatGPT can be prone to unintentionally memorizing and reproducing substantial portions of their training data~\cite{Brown22}, making it notorious for exposing potentially private information within the data~\cite{Carlini21}. Moreover, the vulnerability of ChatGPT to deliberate attacks, that lead to the exposure of its training data and associated personal data, poses an additional threat to user privacy~\cite{WuX23}. This risk is particularly alarming when ChatGPT is used in a work context, as it jeopardizes not only the user's personal data but also the sensitive information of third parties, such as patients or customers. It is important to acknowledge, however, that there are measures in place that prevent certain LLM-based chatbots, such as OpenAI's \say{GPT Team} and \say{GPT Enterprise}, from using user data for training purposes. These models explicitly state that user data is excluded from training by default and offer customizable data retention policies \cite{OpenAI2025Pricing,OpenAI2025EnterprisePrivacy}. While these measures address privacy concerns, this study does not consider such distinctions. At the time of data collection, the market penetration of ChatGPT in professional settings-especially through such subscription-based models-, was considered limited, as the enterprise plan had been available for only five months, and the team plan had launched less than a month prior \cite{OpenAI2025ReleaseNotes}. Thus, the specific version or model of ChatGPT used by participants at that time was not recorded in this study. Individuals who incorporate ChatGPT in their work routines, might not necessarily be aware of those risks~\cite{Kimbel24}. Supporting this, \cite{Ma25} show that both users and creators of custom Generative Pre-trained Transformers (GPTs) often lack clear mental models of how data flows through these systems, revealing gaps between user understanding and official documentation. While previous studies investigated individuals’ perceived benefits and challenges when using ChatGPT at their schools or workplaces~\cite{shoufan2023exploringStudentsPerception} only few have assessed how these benefits align with users' privacy perceptions. This study attempts to address this gap by exploring how individuals perceive privacy risks when using ChatGPT to facilitate work-related tasks. We consider three aspects, which may impact individuals' ChatGPT usage behavior due to privacy considerations: (1) Policies for using GenAI tools employed by the organizations individuals work in, (2) individuals' general privacy concerns in the digital space and (3) individuals' explicit knowledge about how ChatGPT stores and processes data.  We refer to these three aspects as (1) \textit{organizational policies}, (2) \textit{integrated privacy concerns}, and (3) \textit{ChatGPT proficiency}. While investigating how these aspects impact ChatGPT usage behavior and interact with each other, we raise the following research questions:

\par\bigskip 
\begin{description}
    \item[\textbf{RQ1.}] What effect do \textit{organizational policies} have on employees' \textit{ChatGPT proficiency}? \par\bigskip
    
    \item[\textbf{RQ2.}] How do \textit{organizational policies}, individuals' \textit{integrated privacy concerns}, and \textit{ChatGPT proficiency} affect individuals' ChatGPT usage behavior in a work context? \par\bigskip
    
    \item[\textbf{RQ3.}] How does the influence of \textit{integrated privacy concerns} and \textit{ChatGPT proficiency} on ChatGPT usage behavior (as examined in RQ2) differ between individuals working in organizations with established \textit{organizational policies} and individuals working in organizations without such policies? \par\bigskip
    
    \item[\textbf{RQ4.}] How do different types of \textit{organizational policies} affect individuals' ChatGPT usage behavior in a work context?
\end{description}

\textit{Contribution.} To address these questions, we conducted an online survey with $N=224$ participants working in different industry sectors in Europe who use ChatGPT in a work context. We assessed whether \textit{organizational policies}, such as restrictions on GenAI tools or awareness measures, were present in participants' work environment. Furthermore, to assess participants' \textit{integrated privacy concerns}, we introduced a novel construct that integrates key elements from established frameworks for capturing privacy concerns and risks. Moreover, to evaluate participants' \textit{ChatGPT proficiency}, we developed a new construct specifically tailored to this purpose. Unlike established constructs that assess general technological adoption competence, our approach is based on explicit privacy policies of OpenAI \cite{OpenAIPrivacy2023} and extends prior work by quantitatively assessing privacy-related proficiency in the use of ChatGPT in professional settings.

Our findings indicate that organizational policies significantly enhance ChatGPT proficiency, suggesting that structured guidance improves users' understanding of data storage and processing. In addition, organizational policies may encourage individuals to reflect on the utilization of ChatGPT and, as a consequence, gain more insight into how ChatGPT stores and processes data (RQ1). Participants in organizations with such policies also used ChatGPT for a broader range of applications, demonstrating that these guidelines foster more versatile use (RQ2). Privacy concerns were found to negatively affect both the frequency of use and the diversity of use cases, particularly among  users in organizations without specific policies, where individuals relied on their personal privacy concerns and knowledge to guide their behavior (RQ2, RQ3). In contrast, no significant effects of ChatGPT proficiency on ChatGPT usage behavior were observed in organizations with established policies (RQ3). There was no indication that different types of organizational policies differed in their effect on individuals' ChatGPT usage behavior  (RQ4).

\section{Theoretical Background}
\label{Theoretical Background}

This section gives a brief account of the background of LLMs and GenAI tools, their technological adoption in ChatGPT and privacy risks associated with ChatGPT.

\subsection{Large Language Models and GenAI} 
\label{Large Language Model unveiled}

Language Modeling (LM) is a fundamental component in the field of Natural Language Processing (NLP) used in AI~\cite{Zhao23}. LM models play a crucial role in text generation and representation by predicting sequences of words~\cite{Brown20}. These models are designed to estimate the probability of token sequences based on their co-occurrence in the training data~\cite{Brown22}. Their primary aim is to predict the likelihood of future or missing tokens~\cite{Zhao23}. Here, a token can be a character sequence, word, or sub-word, which serves as a semantic unit for processing~\cite{Brown22}.

Scaling these models, whether by increasing the model size or expanding the training data~\cite{Zhao23}, leads to significant enhancements in their performance and capacity~\cite{Han2021}. Such large-scale models are known as LLMs. LLMs are characterized by their vast number of parameters and are trained on extensive corpora of publicly available human-generated text~\cite{Min23}, producing a statistical distribution of tokens~\cite{Shanahan23}. LLMs are applied in various contexts where NLP is required~\cite{Wang23_B}. The advancement of these models has given rise to GenAI, which refers to systems capable of creating new content, such as text, images, and audio-based on patterns learned from data. GenAI, powered by LLMs, is widely utilized in applications that involve natural language generation and understanding, including voice assistants like Amazon Alexa and communication-assisting chatbots~\cite{Brown22}.

\subsection{ChatGPT} 
\label{Technological Adoption: ChatGPT}

ChatGPT, adapted from the GPT series for dialogue purposes to generate human-like responses, has today emerged as the most widely used GenAI tool~\cite{Zhao23}. Models of the GPT series utilize the transformer architecture introduced by \cite{Vaswani23} and are trained on extensive text corpora to learn language structures, semantics, and syntax through unsupervised learning~\cite{Radford18, Zhou23, WuT23}.

ChatGPT, especially since the introduction of GPT-3.5~\cite{Wang23}, employs advanced deep learning techniques to generate responses to input prompts~\cite{Okey23, Eke23}. The GPT series has evolved significantly, from GPT-1 to the latest GPT-4, through iterative improvements~\cite{WuT23, WuX23}. GPT-1, launched in 2018, was a pioneering LLM that used the decoder component of the transformer architecture. It employed a two-step training process: unsupervised pretraining on large, unlabeled text to build a foundational model, followed by supervised fine-tuning with annotated data for specific tasks. GPT-1's strength lay in its ability to learn effectively without extensive labeled datasets, though some labeled data was still required for fine-tuning. GPT-2 advanced the field by increasing the number of parameters and focusing on multitask learning, which reduced reliance on labeled data. This approach enhanced generalizability and performance while maintaining GPT-1’s network structure~\cite{Guo23}. GPT-3, introduced in 2020, further expanded the model's scale by incorporating meta-learning and in-context learning, thus eliminating the need for additional fine-tuning. It also used more parameters and larger datasets during training, and was designed to predict outputs based on queries and examples. Following GPT-3, OpenAI released InstructorGPT, a derivative version of the GPT-3.5 series. This version introduced a three-stage training process: supervised fine-tuning, training a reward model, and reinforcement learning based on this reward model. InstructorGPT served as the prototype for ChatGPT, which employs similar methods with slight variations in data collection~\cite{Okey23}. The latest iteration, GPT-4, released in 2023, showcases enhanced generation capabilities and supports both textual and visual inputs.

Along with its widespread acceptance and application in private settings, ChatGPT demonstrates significant potential for work-related tasks across various fields~\cite{Akbar23}. Its utilization is already being discussed for academic and educational settings~\cite{Memarian23, Muneer23}, as well as in research and broader work environments~\cite{Dis23, Sandoval23}. Notably, previous work has already investigated the chatbot's potential in various domains, such as education~\cite{Park2024_ChatGPTEducation,Han2024_ChatGPTElementary}, healthcare~\cite{Calle2024_AIDrivenHealthcare}, and IT~\cite{Kabir2024_ChatGPTStackOverflow}. These investigations highlight ChatGPT's utility in streamlining tasks, facilitating workflows, and thereby enhancing overall efficiency. 

\begin{figure*}[!t]
\centering
    \IfFileExists{images/ChatGPT.pdf}{%
        \includegraphics[width=0.95\linewidth]{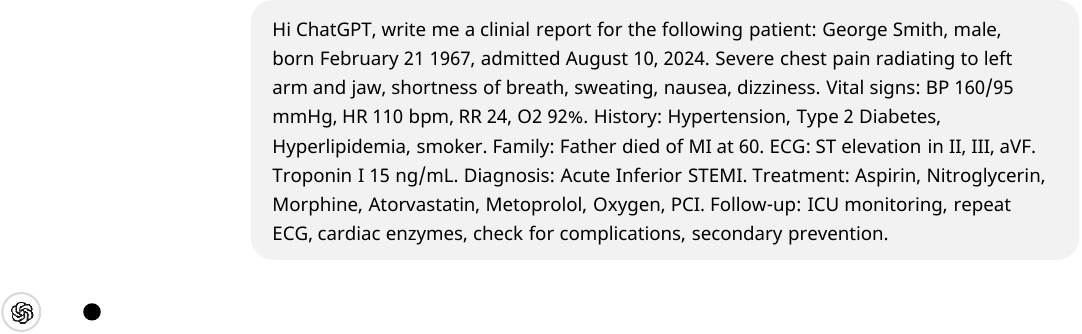}%
    }{%
        \missingfigure{images/ChatGPT.pdf}%
    }
    \caption{A fictional user input disclosing confidential patient data in OpenAI's chatbot ChatGPT to facilitate creating a clinical report: Using ChatGPT to write clinical reports more efficiently, medical staff might (unintentionally) disclose highly sensitive health data without the patient's consent. \authorssource}
    \label{fig:userprompt}
\end{figure*}

\subsection{Privacy Challenges of ChatGPT}
\label{Privacy Challenges of ChatGPT}

While ChatGPT offers indisputable benefits in completing work-related tasks, using it also introduces certain privacy risks that users might not be fully aware of~\cite{Kimbel24}. For example, ChatGPT not only processes personal data linked to a user’s account, but also stores user conversations. OpenAI uses this conversation data, which may include sensitive or personal details, to further train their models~\cite{WuX23}. Incorporating user inputs into their training datasets poses the risk of inadvertent data exposure to unauthorized parties~\cite{Brown22}. Furthermore, OpenAI retains the authority to internally process this conversation data, e.g., reviewing user prompts for potential compliance violations~\cite{OpenAI_PrivacyPolicy}. This practice can additionally lead to the potential exposure of user data to OpenAI, compromising the confidentiality of the information shared. Additionally, there have also been instances of users' unintentionally disclosing sensitive information. As a recent example, ChatGPT has been banned in Australian hospitals after medical staff in several hospitals prompted patient data into ChatGPT to write clinical reports without the patients’ consent~\cite{Guardian2023_AustralianHospitals}. Fig.~\ref{fig:userprompt} shows a fictional user prompt of such type of data disclosure. While creating a clinical report with ChatGPT certainly saves time and can improve the reports' quality, doing so without the patient's consent constitutes a violation of their privacy. Beyond these deliberate practices, user data can also be compromised by technical errors or targeted attacks from malicious actors~\cite{King24}. In 2023, a security vulnerability in an open-source library used by OpenAI became public~\cite{OpenAIChatGPTOutage}. This led to the exposure of users' chat titles, names, and banking details to other active users. As another example, researchers have demonstrated how specific input prompts can be leveraged to generate output that reveals parts of the training data set that contain confidential user data~\cite{Nasr2023_PoemForever}. These privacy concerns are particularly critical in a work context, where using ChatGPT with internal information can lead to unintended exposure of sensitive data and potentially compromise the confidentiality of third parties involved, including information stored in customer databases and health records. More critically, there is a risk of revealing data about uninvolved third parties, as this data is stored in customer databases, health records, and other confidential repositories.

\section{Related Work} 
\label{sec:related_work}

This section outlines prior research on user behavior in adopting GenAI tools, emphasizing the relationship between users' privacy perceptions and technology adoption.

Recent studies have systematically examined LLMs from a privacy perspective, addressing concerns and risks tied to their use, potential mitigation strategies, and frameworks for evaluating their suitability in privacy-sensitive contexts. \cite{Kibriya24} and \cite{Smith23} highlight risks across both training and usage phases, such as unintentionally memorizing and reproducing parts of their training data or the inability of LLMs to unlearn or selectively remove specific information. Adding to this, \cite{Shahriar25_A} provide a comprehensive review of privacy-enhancing technologies for text data, categorizing common risks and defense strategies. Complementing these efforts, they also propose a framework for evaluating LLMs’ privacy-awareness and competencies for their use in realistic scenarios.

Given the growing understanding of LLMs’ privacy challenges, it is increasingly important to understand how these issues manifest in end-user applications. Chatbot technology, which leverages AI’s natural language capabilities, has emerged as a prominent application area, rapidly gaining popularity due to its interactive and conversational nature. \cite{Cheng20} highlight a gap in prior research: the impact of perceived privacy risks and concerns about disclosing personal information on users' satisfaction and technology adoption. To address this gap, \cite{Eeuwen17} and \cite{Rese20} conducted online surveys to explore the acceptance and intention to use AI chatbots, with a focus on privacy considerations. Both studies extended the Technology Acceptance Model (TAM) to investigate how privacy concerns influence users' attitudes and intentions to adopt technology. \cite{Eeuwen17} propose to integrate TAM with Innovation Diffusion Theory to assess the effect of privacy concerns on users' attitudes toward technology adoption. Similarly, \cite{Rese20} apply TAM and Uses and Gratifications Theory to evaluate the acceptance of chatbot technology, considering privacy issues. 

Building upon TAM and its derivative models, these studies examined users' willingness to accept AI technology and their intentions to use it. TAM and its derivatives (e.g., perceived usefulness and perceived ease of use) provide a framework for understanding user acceptance by examining how various factors influence initial attitudes toward technology adoption. By integrating privacy-related theories and factors into these foundational models, the studies investigated how privacy concerns might influence adoption behavior. However, TAM is primarily designed to explain initial adoption behavior and is limited in its ability to capture post-adoption dynamics, which are more strongly shaped by emotional, trust-related, or risk-averse considerations, especially in relation to privacy. Since our study focuses on how privacy-related aspects impact the usage behavior and usage competence of established users rather than their initial intention to use the technology, the TAM is not suitable for our research. Modifying TAM by removing redundant components and adding aspects tailored to our research needs would involve a significant restructuring of the original framework. Instead, we opted to develop a new model tailored to established users. This approach also is more fitting for our research, as it facilitates the integration of a novel method for evaluating user proficiency. Furthermore, some studies specifically investigated users' security and privacy concerns regarding AI-based chatbots. \cite{Saglam21} explored users' concerns about the use of sensitive data by AI chatbots. Their study focused on understanding participants' concerns about AI chatbot interactions and the factors influencing their trust in these systems and impacting technology adoption.

Building on earlier research on AI-powered chatbot technology, recent studies have shifted their focus to advanced LLM-based conversational agents, such as ChatGPT, exploring privacy, trust, and adoption behaviors in this domain. \cite{Glorin23} examined the general usage behavior concerning AI chatbot technology, focusing on perceived cybersecurity risks and privacy concerns related to ChatGPT and similar chatbots, through an online survey. \cite{Choudhury23} built on the work of \cite{Glorin23} by studying the impact of trust on ChatGPT adoption, emphasizing how trust affects both technological adoption and user acceptance. They conducted a web-based survey to assess users' attitudes toward trust-related constructs. \cite{Alkamli24} examined privacy concerns related to ChatGPT using a data-driven approach that combined Twitter data analysis with a user survey, identifying privacy leakage from public data exploitation, personal input exploitation, and unauthorized access as the three main areas of concern among users. Expanding on this research, \cite{Menon23} investigated factors influencing users' acceptance of LLM-based chatbot technology, with a particular focus on ChatGPT. They applied the Unified Theory of Acceptance and Use of Technology framework, incorporating constructs related to privacy risks, concerns, and trust. While these studies provided first insights into how users in general perceive security and privacy risks of GenAI tools such as ChatGPT, they have not examined the impact of privacy concerns on the usage behavior in work-related settings. Furthermore, none of these studies have investigated the impact of privacy concerns on users' technological competence. 

Addressing this in their work, \cite{Kimbel24} conducted a qualitative study using semi-structured interviews to explore ChatGPT adoption in work settings. Their research focused on organizational policies, user awareness of security and privacy risks, and strategies for mitigating these risks. However, due to the study's qualitative nature, it primarily offers insights based on anecdotal feedback from participants. To the best of our knowledge, our study is the first to provide quantitative insights into the privacy-driven behavior of ChatGPT users in a work context. Expanding on this, \cite{Zhang24} conducted a qualitative analysis combining real-world ChatGPT interaction histories with semi-structured interviews to investigate how users adapt their usage behavior to mitigate disclosure risks when engaging with LLM-based conversational agents. The sample analysis revealed that users disclose a wide spectrum of information, ranging from less directly identifiable details to highly sensitive personal data, both about themselves and others. Additionally, the analysis of the usage samples and interviews identified various privacy-protective strategies employed by users, such as anonymizing or falsifying information, to manage risk associated with the disclosure of personal identifiable data. The study thereby provides a comprehensive overview of actual end-user disclosure behavior when using ChatGPT.

Turning attention to platform customization, \cite{Ma25} investigated privacy perceptions of custom GPTs on OpenAI's platform using semi-structured interviews. In addition to users' privacy concerns and enacted practices, their work examined both users' and creators' mental models of how data flows within these systems. Their analysis compared participants’ conceptions of three GPT usage scenarios against OpenAI’s official privacy-policy documentation, revealing widespread uncertainty about how data actually flows. Importantly, while both \cite{Ma25} and our work share an interest in users' policy-oriented understandings, \cite{Ma25} focused on custom GPTs and used qualitative interviews to surface perceptions and practices. By contrast, our study quantitatively assessed users’ proficiency with privacy concepts using questions derived directly from OpenAI's official policies and examines behavior in a work-related context, allowing a complementary comparison between both approaches.

While prior research on individuals' privacy perceptions regarding GenAI tools examined the impact of privacy perceptions on adoption and usage behavior, there has been limited exploration of how these perceptions may stem from users' lack of understanding and competence in handling these technologies. Additionally, the potential influence of the privacy paradox - where users express strong privacy concerns but engage in contradictory behaviors - has not been adequately considered. However, these phenomena have been explored in related domains. For instance, \cite{Netter13} and \cite{Cetto14} examined the gap between users' actual privacy settings on online social networks like Facebook and their perceived or intended settings. The findings of \cite{Netter13} suggest that while users are generally aware of their privacy preferences and favor more restrictive settings, many lack the competence to manage them effectively. This mismatch underscores broader issues of user awareness and control, which could similarly arise in the context of GenAI tools.

\section{Constructs and Research Questions}
\label{sec:Definition of constructs}

The following section outlines the constructs we employed in our study and the research questions we addressed. An overview of the constructs is given in Table~\ref{tab:constructs}. The full questionnaire with the measures used for each construct is included in Appendix~\ref{sec:Questionnaire}.

\subsection{Definition of Constructs}

\paragraph{Organizational policies.} We aimed to assess whether the organizations where participants used ChatGPT had established policies regarding the use of GenAI tools. These policies could range from restricting access to GenAI tools, permitting their use only for specific tasks, to implementing general awareness initiatives that inform employees about potential risks associated with using GenAI tools like ChatGPT. \par\medskip

\begin{table*} [t]
\footnotesize
  \caption{Overview of the constructs \textit{organizational policies}, \textit{integrated privacy concerns}, \textit{ChatGPT proficiency}, and ChatGPT usage behavior (i.e., \textit{frequency of use} and \textit{use cases}). \authorssource}
  \label{tab:constructs}
 \begin{tabular}{p{4.5cm}>{\raggedright\arraybackslash}p{7.5cm}p{2.5cm}}
    \toprule
    Construct & Definition & Reference\\
    \midrule
    \textbf{Organizational policies} & Organizational guidelines and rules that govern the use of generative AI tools by employees, including restrictions on access, permissible use cases, and measures to raise awareness of potential privacy and security risks. & self-defined  \\ \\
     \textbf{Integrated privacy concerns} & Subjective perception of risks and concerns associated with potential disclosure of personal information to third-party entities on the internet, encompassing data-related dimensions including data collection, unauthorized access, and unauthorized secondary use. & \cite{Dinev06_B}, \cite{Smith11}, \cite{Smith96}  \\ \\
  
     \textbf{ChatGPT proficiency} & Explicit knowledge on data processing and storage of ChatGPT & ChatGPT privacy policy~\cite{OpenAIPrivacy2023} \\ \\

  \textbf{Frequency of use} & The frequency participants use ChatGPT in their work context. & self-defined \\ \\

     \textbf{Use cases}  & Amount of different uses cases respectively applications participants use ChatGPT for. & self-defined \\ \\
    
    \bottomrule
  \end{tabular}
\end{table*}

\paragraph{Integrated privacy concerns.} With the construct \textit{integrated privacy concerns}, we want to capture an individual's subjective perception of risks and concerns associated with potential losses linked to users disclosing personal information to third-party entities on the internet. Due to the absence of a broadly agreed-upon definition of privacy~\cite{Dinev06_A} and the near impossibility of measuring privacy itself~\cite{Smith96}, empirical privacy research in the social sciences tends to measure individuals' self-reported \textit{privacy concerns} instead~\cite{Smith11}. \cite{Dinev06_B} define privacy concerns as concerns about opportunistic behavior related to the personal information submitted by internet users. These concerns mainly revolve around the collection of data, data errors, unauthorized access, and unauthorized secondary use of the information collected. These dimensions have since served as some of the most reliable scales for measuring individuals' concerns toward organizational privacy practices~\cite{Smith11}. Being a multidimensional construct, users' privacy concerns, however, are frail to influence and be influenced by other constructs~\cite{Malhotra04, Smith96}. Given the extensive terminology used to define privacy concerns and their influencing factors, it is difficult to adopt a single definition verbatim, as the specific wording of these definitions does not fully match the scope of our study. ~\cite{Smith96} have previously affirmed this through their examination of various dimensions underlying information privacy concerns. They stated that the specific dimensions differ from study to study, making it nearly impossible for a common, unifying framework to emerge. As previously stated, our primary focus is not on the further development, application, and evaluation of these constructs nor the creation of new ones to capture the general attitude toward privacy in the context of technology adoption. Instead, our emphasis is on exploring whether users' general attitude toward privacy can be seen explicitly in the utilization of ChatGPT. Consequently, we have incorporated the established constructs by \cite{Smith11}, \cite{Smith96} and ~\cite{Dinev06_B} into our work in a slightly modified form. Given the interconnected relationship between the constructs of perceived privacy concerns and perceived privacy risks, we chose to represent those within one unifying construct named \textit{integrated privacy concerns}, representing a framework consolidating the substantial elements of both constructs. 

\paragraph{ChatGPT proficiency.} This variable refers to the proficiency in the practical utilization of ChatGPT, guided by a thorough understanding of its principles and policies. Established constructs for evaluating technological proficiency primarily aim to capture general competence within a broader sense. In this study, we want to assess the privacy-related proficiency of a particular tool, namely ChatGPT. As described in the introduction to this paper, this is based on the understanding that proficiency in using ChatGPT is considered a prerequisite for responsible and ethically sound behavior. The validated constructs designed for measuring technological proficiency can, therefore, not be directly applied to this particular case. Consequently, we established a new construct labeled \textit{ChatGPT proficiency}. This construct is directly derived from OpenAI's privacy policies and the technical functionalities of ChatGPT~\cite{OpenAI_PrivacyPolicy}, thereby specifically tailored to assess users' competences in applying ChatGPT. 

\paragraph{Frequency of use.} As the first aspect of ChatGPT usage behavior, \textit{frequency of use} relates to how regularly participants use ChatGPT in their work context.

\paragraph{Use cases.} To capture ChatGPT usage behavior from a more qualitative, content-related perspective, \textit{use cases} specifies the application purposes for which individuals use ChatGPT in their work context, e.g, data analysis or language translation. In particular, we focus on how many different use cases, on average, participants use ChatGPT for.

\subsection{Research Questions}
\label{sec:rqs}

\paragraph{Organizational policies and ChatGPT proficiency.} In our research, we aim to explore individuals' perceptions of privacy regarding ChatGPT, as well as how organizational policies impact these perceptions. Our first research question examines the extent to which \textit{organizational policies} affect employees' proficiency in using ChatGPT. Furthermore, we want to investigate whether employees in organizations with defined \textit{organizational policies} are more informed about ChatGPT's data access, storage, and processing practices, as well as the associated liability implications (i.e., \textit{ChatGPT proficiency}), compared to employees in organizations without such policies. 

\begin{description}
\item[\textbf{RQ1}]What effect do \textit{organizational policies} have on employees' \textit{ChatGPT proficiency}?
\end{description} 

\paragraph{ChatGPT usage behavior.} Furthermore, we investigate how \textit{organizational policies}, \textit{ChatGPT proficiency} and \textit{integrated privacy concerns} impact the ChatGPT usage behavior of individuals, i.e.,  how regularly (\textit{frequency of use}) and for which variety of applications (\textit{use cases}) they use ChatGPT. That is, we aim to find out how individuals' ChatGPT usage behavior is affected (1) by them experiencing organizational policies regarding GenAI tools, (2) by their general concern for online privacy (i.e., \textit{integrated privacy concerns}), and (3) by the degree to which they are informed about potential privacy risks (i.e., \textit{ChatGPT proficiency}). 

\begin{description}
\item[\textbf{RQ2.}]How do (1) \textit{organizational policies}, (2) individuals' \textit{integrated privacy concerns}, (3) \textit{ChatGPT proficiency} affect individuals' ChatGPT usage behavior in a work context?
\end{description}

\paragraph{Group comparison of organizations with and without organizational policies.} On this basis, we aim to explore whether the factors potentially influencing individuals' ChatGPT usage behavior (as addressed in RQ2) vary between organizations with \textit{organizational policies} related to GenAI and organizations without such policies. Specifically, we want to understand how the impact of \textit{integrated privacy concerns} and \textit{ChatGPT proficiency} differs depending on the presence or absence of these organizational policies. In other words, we are interested in determining whether the organizational context—defined by the presence or absence of specific policies—affects how privacy concerns and proficiency shape individuals' use of ChatGPT: 

\begin{description}
\item[\textbf{RQ3.}] How does the influence of \textit{integrated privacy concerns} and \textit{ChatGPT proficiency} on ChatGPT usage behavior (as examined in RQ2) differ between individuals working in organizations with established \textit{organizational policies} and individuals working in organizations without such policies?
\end{description} 

\paragraph{Comparison of different types of organizational policies.}
Lastly, we want to give a deeper insight into how different kinds of organizational policies, namely \textit{usage constraints} and \textit{user constraints}, influence individuals' usage behavior: 

\begin{description}
\item[\textbf{RQ4.}] How do different types of \textit{organizational policies} affect individuals' ChatGPT usage behavior in a work context?
\end{description}

\section{Method}
\label{Method}

The following section describes the data collection and validation process, along with the analytical methods used during the data analysis phase.

\subsection{Sample and Data Collection Procedures}
\label{Data Collection and Validation}

We addressed our research questions in an online survey among participants who regularly use ChatGPT in their work context. Regarding background variables, we gathered information on participants' \textit{gender}, \textit{age}, and their highest level of (formal) \textit{education}. For demographic details, refer to Table~\ref{tab:demographics}. Details on the definition of the groups for the individual background variables can be found in the appendix~\ref{sec:Questionnaire}. An additional factor that we did not include in our analyses, but which we collected in order to better understand the sample, is the participants' sector of occupation (see Table~\ref{tab:occupation}). Participants represented a wide range of sectors, with the IT \& technology industry being the most prevalent. This broad sector representation allows our study to offer diverse perspectives on ChatGPT usage across various industries. 

Prior to distribution, the questionnaire was validated through a small pilot study involving $N=11$ participants to identify and address incomprehensibility, inconsistencies, and language-related problems. For participant recruitment, we employed a mixed strategy, using both the crowdsourcing platform Prolific Academic and targeted outreach via email. The questionnaire itself was hosted on Google Forms. All participants were redirected to the questionnaire using a one-time-use web link.


\begin{table}[h]
\footnotesize
\caption{Demographic characteristics of the participants. \authorssource}
\label{tab:demographics}
\centering
\begin{tabular}{p{6cm} l}
\toprule
\textbf{Characteristics (N=224)} & \textbf{n}  \\
\midrule
\textbf{Gender} &  \\
\quad Male & 133  \\
\quad Female & 89  \\
\quad Non-binary & 2  \\
\midrule
\textbf{Age}&  \\
\quad 18–24 Years & 69 \\
\quad 25–34 Years & 105  \\
\quad 35–44 Years & 32  \\
\quad 45+ Years & 18  \\
\midrule
\textbf{Level of (formal) education} & \\
\quad Foundational education& 49\\
\quad Advanced education& 14   \\
\quad Academic education& 161  \\
\bottomrule
\end{tabular}
\end{table}%

\begin{table}[h]
\footnotesize
\caption{Employment sectors of the participants. \authorssource}
\label{tab:occupation}
\centering
\begin{tabular}{p{5.5cm} l}
\toprule
\textbf{Sector} & \textbf{n} \\
\midrule
IT \& technology & 75 \\
Engineering \& manufacturing & 26 \\
Education \& research & 25 \\
Finance \& banking & 12 \\
Business \& consulting & 15 \\
Healthcare \& pharmaceuticals & 14 \\
Arts \& entertainment & 13 \\
Other & 44 \\
\bottomrule
\end{tabular}
\end{table}

To ensure the quality of responses, two attention checks were incorporated. Users who incorrectly answered one of the attention check questions were later excluded from further consideration. Additionally, participants were not allowed to submit the survey with incomplete responses. At the beginning of each section, we provided brief instructions addressing the answers to the respective sets of questions. An additional criterion employed in the categorization of received questionnaire responses involved identifying thematic inconsistencies during the analysis process. Responses falling under this criterion exhibited conflicting evaluations of thematically analogous and interconnected statements and were consequently omitted from consideration.

\subsection{Measures}
\label{Measures}

In the following, we describe the measures employed for our survey. The complete questionnaire can be found in Appendix~\ref{sec:Questionnaire}.

\paragraph{Organizational policies.} To better understand not only whether \textit{organizational policies} regarding the use of GenAI were present in the participants' organizations but also which kind of policies these were, participants were presented with eight different options (e.g., \textit{training and awareness measures} or \textit{usage restricted for documents of certain confidentiality classification}) to choose from in a multiple-response format.  For further analysis (RQ4), items 3 to 6 were categorized as \textit{usage constraints}, indicating that the policy referred to \textit{how} employees were supposed to use a GenAI tools and items 1 and 2 as \textit{user constraints}, indicating that the policy referred to \textit{who} was allowed to use particular GenAI tools. 

\paragraph{Integrated privacy concerns.} Participants' \textit{integrated privacy concerns} were measured by having them self-assess their attitudes and concerns regarding privacy through six items (e.g., \say{Sharing my personal information does not bother me, so I usually don't hesitate to provide it.}). These items were rated on a five-point Likert scale, where 1 indicated "strongly agree" and 5 indicated "strongly disagree." For later analysis, the scale was reversed so that higher scores represent a stricter, more concerned attitude toward privacy. To validate the scale, we conducted an exploratory factor analysis using principal axis factoring with promax rotation. The analysis revealed that all items loaded onto a single factor as expected. In addition, the  internal consistency of the items was satisfactory ($\alpha = .86$).

\paragraph{ChatGPT proficiency.} Participants' \textit{ChatGPT proficiency} was assessed using eleven knowledge-based items presented as questions in a \say{true/false} format. These questions were designed to evaluate participants' understanding of ChatGPT's privacy policies and technical functionalities (e.g., \say{ChatGPT's privacy policy guarantees that personal information is anonymized.}). Content validity was secured by directly sourcing each question from OpenAI's publicly available security and privacy policies. In addition, two of the authors iteratively reviewed and refined the questions to ensure clarity and alignment with the policies. To prevent participants from only guessing the correct answer, they could also choose \say{unsure} as an option, which was treated as an incorrect answer. The \textit{ChatGPT proficiency} score for each participant was calculated as the percentage of questions they answered correctly. 

\paragraph{Frequency of use.} Participants indicated how often they use ChatGPT in their work context on a four-point Likert scale ranging from 1 ("less than once a week") to 4 ("daily usage"). 

\paragraph{Use cases.} To assess the \textit{use cases} participants employed ChatGPT for, participants could choose from a list of eight types of purposes in a multiple-response format (e.g., \say{Email-Assistance: In my workflow, ChatGPT is employed for automatically summarizing emails and crafting responses to the initial message.}).

\begin{table*}[t!]
\small
\centering
\begin{threeparttable}

\caption{Types of \textit{organizational policies} regarding GenAI tools and \textit{use cases} of ChatGPT usage, with their respective frequencies. \authorssource}
\label{tab:combined_policies_usecases}
\begin{tabular}{p{13cm} l}
\toprule

\textbf{Type of Organizational Policy} & \textbf{n}  \\
\midrule 
\quad De-identification or anonymization of input data for ChatGPT & 47 \\
\quad Explicit prohibition of including personal information in input data & 43 \\
\quad Limiting usage to documents classified up to a specified level of confidentiality & 41 \\
\quad Restricting sharing or external use of content generated by ChatGPT & 36 \\
\quad Restricting usage to specific departments or circle of users & 33 \\
\quad Requirement of prior training or awareness-raising measures for authorization & 12 \\
\quad Other unspecified restrictions & 13 \\
\midrule
\textbf{Use Case} & \textbf{n} \\
\midrule 
\quad Brainstorming and idea assistance & 141 \\
\quad Language translation \& communication assistance & 111 \\
\quad Content polishing & 83 \\
\quad E-mail assistance & 74 \\
\quad Code assistance & 74 \\
\quad Data analysis assistance & 46 \\
\quad Content generation & 43 \\
\quad Decision-making assistance & 41 \\
\quad Support assistance & 35 \\
\quad Meeting assistance & 15 \\

\bottomrule
\end{tabular}
\begin{tablenotes}
\textit{Note: Participants were allowed to select multiple responses for both constructs.}
\end{tablenotes}
\end{threeparttable}
\end{table*}

\subsection{Analyses}
\label{analyses}

To address our research questions, we first conducted descriptive analyses, i.e., we inspected means and standard deviations for the study variables. We then used \textit{t}-tests to assess the bivariate effects of the categorical background variables (i.e., \textit{gender}, \textit{age}, and \textit{education}) on endogenous variables (i.e., \textit{integrated privacy concerns}, \textit{ChatGPT proficiency}, \textit{frequency of use}, and \textit{use cases}). For these tests, we utilized dummy variables to examine the effects of individual categories of the background variables. Next, we performed correlation analyses for all study variables. Specifically, three sets of correlation analyses were executed: one for the entire cohort and group-specific correlation analyses for participants with and without \textit{organizational policies}. Finally, to simultaneously examine both direct and indirect effects outlined in our research questions, we employed path modeling using Mplus 8~\cite{Muthen2010_mplus}. This analysis involved five distinct models: one encompassing all study variables across all participants (addressing RQ1 and RQ2), and two models exploring the effects of \textit{integrated privacy concerns} and \textit{ChatGPT proficiency} on \textit{frequency of use} and \textit{use cases} for participants with and without \textit{organizational policies} (addressing RQ3). Furthermore, we spcified two models to examine if there was a difference between the two main types of \textit{organizational policies} (i.e., \textit{user constraints} and \textit{usage constraints}) within the group of employees in organizations with policies (adressing RQ4). We controlled for relevant background variables that significantly contributed to the models. To evaluate model fit, established fit indices and cut-off criteria (i.e., 
$0 \le \chi^2/df \le 2$, 
$0.97 \le \mathrm{CFI} \le 1.00$, 
$0.97 \le \mathrm{NNFI} \le 1.00$, 
$0 \le \mathrm{RMSEA} \le 0.05$, 
$0 \le \mathrm{SRMR} \le 0.05$) were employed (see Figures~2, 3, and~4). As two participants had missing values for \textit{education}, these cases were excluded from further analyses, reducing the sample size to $N=224$. The full dataset, along with the analysis syntax for both SPSS and Mplus 8, is available as supplementary material to this paper.

\subsection{Ethics}
\label{Ethics}

The user study was designed in accordance with the guidelines of the ethics committee of the first author's institution. Before participants started the questionnaire, we asked for their explicit consent for data collection, emphasizing voluntary participation with the freedom to withdraw at any stage. Acknowledging the potential sensitivity of some questions, we included optional responses, allowing participants to abstain from answering items they perceived as uncomfortable or offensive. Participants were asked for their Prolific IDs solely to match their questionnaires with recruited accounts from Prolific for compensation. The IDs were deleted afterward to ensure anonymity.


\section{Results}
\label{Results} 

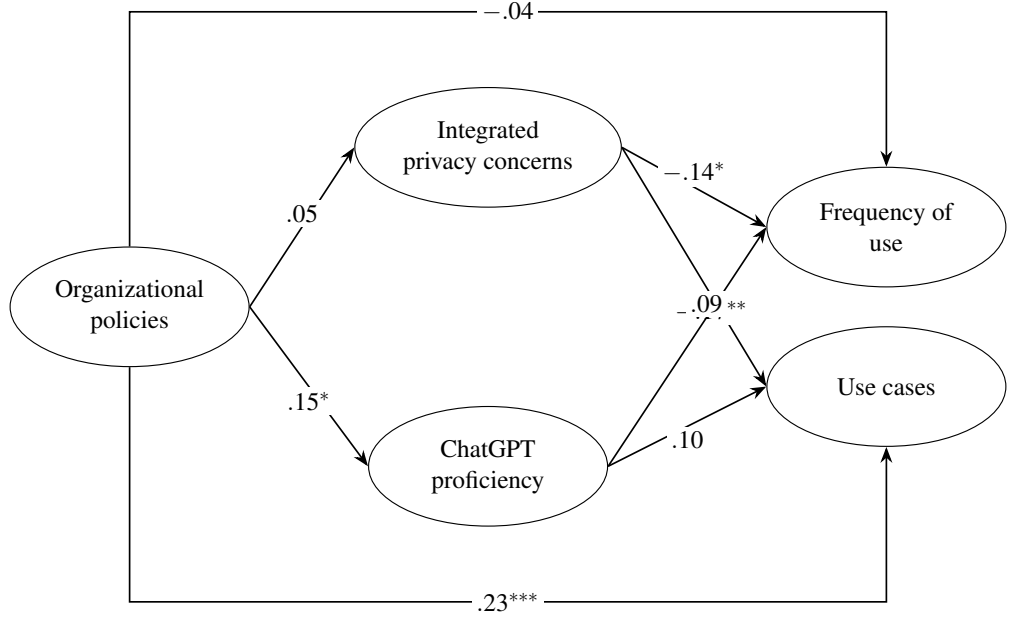
\begin{figure*}[t]
\centering
\resizebox{0.8\textwidth}{!}{%
\begin{tikzpicture}
    \node[sem-variable] (policies) at (0,2)
        {Organizational\\policies};
    \node[sem-variable] (concerns) at (4.5,4)
        {Integrated\\privacy concerns};
    \node[sem-variable] (proficiency) at (4.5,0)
        {ChatGPT\\proficiency};
    \node[sem-variable] (frequency) at (9.5,3)
        {Frequency of\\use};
    \node[sem-variable] (usecases) at (9.5,1)
        {Use cases};

    \draw[sem-path] (policies.east) -- (concerns.west)
        node[sem-coefficient, midway, above] {$.05$};
    \draw[sem-path] (policies.east) -- (proficiency.west)
        node[sem-coefficient, midway, below] {$.15^{*}$};
    \draw[sem-path] (concerns.east) -- (frequency.west)
        node[sem-coefficient, midway, above] {$-.14^{*}$};
    \draw[sem-path] (proficiency.east) -- (usecases.west)
        node[sem-coefficient, midway, below] {$.10$};
    \draw[sem-path] (concerns.east) -- (usecases.west)
        node[sem-coefficient, pos=.62, below] {$-.17^{**}$};
    \draw[sem-path] (proficiency.east) -- (frequency.west)
        node[sem-coefficient, pos=.62, above] {$.09$};

    \draw[sem-path] (policies.north) -- (0,5.7) -| (frequency.north);
    \node[sem-coefficient] at (4.75,5.7) {$-.04$};
    \draw[sem-path] (policies.south) -- (0,-1.7) -| (usecases.south);
    \node[sem-coefficient] at (4.75,-1.7) {$.23^{***}$};
\end{tikzpicture}%
}%
\caption{Path model for RQ1 and RQ2. Effects of background variables are omitted for readability. $N=224$. ${}^{*}p<.05$, ${}^{**}p<.01$, and ${}^{***}p<.001$. Model fit: $\chi^2/df=2.03$, CFI$=.96$, NNFI$=.81$, RMSEA$=.07$, and SRMR$=.03$. \authorssource}
\label{fig:CISEM1}
\end{figure*}

\subsection{Descriptives and Correlations}

Descriptive statistics revealed that 54\% ($n=121$) of participants reported that there were no \textit{organizational policies} regarding GenAI tools usage in their organizations (or they were not aware of those). Among the remaining $n=103$ participants, who reported that such policies existed in their organizations, the most frequently selected option was \say{De-identification or anonymization of input data for ChatGPT}, chosen by 46\% of this group. The average score for \textit{integrated privacy concerns} was $M=3.88$ ($SD = 0.81$), indicating that participants generally have a relatively strict and concerned attitude toward privacy. Conversely, \textit{ChatGPT proficiency} was low, with only 36\% of statements correctly validated ($SD = 18\%$), suggesting that participants generally lack familiarity with and expertise in using ChatGPT.
Two notable misconceptions were observed: First, 90\% of participants mistakenly believed that their personal information would be anonymized (see Appendix~\ref{sec:Questionnaire}, item 9). However, the privacy policies only mention the possibility of aggregating or anonymizing personal information, without providing an explicit guarantee. Second, 84\% incorrectly assumed that OpenAI ensures the security of online communication channels (item 4), which is not explicitly stated in the privacy policies. 

Regarding \textit{frequency of use}, 50 participants (22.3\%) reported using ChatGPT less than once a week, and 47 participants (21.0\%) used it once a week. In contrast, the majority used ChatGPT more frequently, with 94 participants (42.0\%) using it several times a week and 33 participants (14.7\%) using it daily. When examining the number of different \textit{use cases} per participant, it was found that participants generally utilize ChatGPT for multiple applications. The average number of \textit{use cases} per participant was $M=3.00$ ($SD = 1.41$), while only a minority used ChatGPT for just one application (i.e., 29 participants (12.9\%)). A significant portion used it for two different use cases (62 participants, 27.7\%) or three different use cases (68 participants, 30.4\%) as indicated in the questionnaire. Overall, \say{Language translation and communication assistance} emerged as the most common application (49.6\%). An overview of participants' distribution across all use cases is provided in Table~\ref{tab:combined_policies_usecases}. The descriptive statistics for participants from organizations with and without policies in place are presented in Tables~\ref{tab:correlation_policies} and \ref{tab:correlation_no_policies} in the Appendix.

For the entire cohort ($N=224$), Pearson’s $r$ correlations revealed a moderate negative relationship between \textit{integrated privacy concerns} and \textit{frequency of use} ($r = -.14, p < .05$) and between \textit{integrated privacy concerns} and \textit{use cases} ($r = -.16, p < .05$). Additionally, a substantial positive relationship was observed between \textit{frequency of use} and \textit{use cases} ($r = .42, p < .01$). Furthermore, a moderate correlation was found between \textit{organizational policies} and \textit{ChatGPT proficiency} ($r = .15, p < .05$). In addition, a slightly stronger correlation between \textit{organizational policies} and \textit{use cases} ($r = .23, p < .01$) was determined. In the group without organizational policies ($n=121$), correlations were similar to those for the overall cohort, with an additional moderate correlation observed between \textit{ChatGPT proficiency} and \textit{frequency of use} ($r = .20, p < .05$). In the group with organizational policies ($n=103$), significantly fewer correlations were found, with only one significant correlation between \textit{frequency of use} and \textit{use cases} ($r = .39, p < .01$). All correlation values are detailed in Appendix~\ref{sec:Questionnaire}.

Regarding effects of background variables, \textit{t}-tests indicated that participants with a foundational level of \textit{education} reported fewer \textit{integrated privacy concerns} ($t_{edu_{foundational}} = -2.89, p < .01, d = .45 $), whereas those with an academic level of \textit{education} reported more \textit{integrated privacy concerns} ($t_{edu_{academic}} = 2.55, p < .05, d = .37$). Additionally, participants within the highest age group (45+ years) reported a lower \textit{frequency of use} ($t = -2.71, p < .01, d = .69$).

\subsection{Structural Effects}
\label{sec:effects}

This section outlines the key findings of our study in relation to the three research questions proposed above. Using path modeling, we tested our theorized effects while controlling for relevant background variables. In the first model (see Fig.~\ref{fig:CISEM1}), we tested how \textit{organizational policies} affects \textit{ChatGPT proficiency} (RQ1) and which effect \textit{ChatGPT proficiency} and \textit{integrated privacy concerns} has on ChatGPT usage behavior (i.e., \textit{use cases} and \textit{frequency of use}) (RQ2). In addition, it was controlled for effects of the background variable \textit{education}. 

\paragraph{RQ1.} For RQ1, we found a significant effect of \textit{organizational policies} on participants' \textit{ChatGPT proficiency} ($\beta = .15, p < .05$), indicating that participants in organizations with GenAI policies are more knowledgeable about data storage and processing in ChatGPT compared to those in organizations without such policies. 

\paragraph{RQ2.} Regarding RQ2, our model shows that participants' in organizations that have GenAI policies in place, use ChatGPT for a wider range of \textit{use cases} ($\beta = .23, p < .001$). This shows that \textit{organizational policies} do not limit the application of ChatGPT for work-related tasks, but, on the contrary, lead to individuals employing ChatGPT in a more versatile manner. \textit{Integrated privacy concerns} predict both \textit{frequency of use} ($\beta = -.14, p < .05$) and \textit{use cases} ($\beta = -.17, p < .01$). This suggests that participants with greater privacy concerns use ChatGPT significantly less frequently and for fewer types of applications. Conversely, no significant effect was observed for \textit{ChatGPT proficiency}. Regarding background effects, we found a significant effect of \textit{education} on \textit{integrated privacy concerns}, i.e., participants with foundational \textit{education} exhibited less restrictive privacy concerns compared to participants with higher levels of \textit{education} ($\beta = -.19, p < .01$). 

\paragraph{RQ3.} For RQ3, we specified two path models to examine the relationships between \textit{integrated privacy concerns}, \textit{ChatGPT proficiency}, and ChatGPT usage behavior variables for two groups: employees working in organizations with and without \textit{organizational policies} regarding GenAI tools. In the group \textbf{without} \textit{organizational policies}, we observed an effect of \textit{integrated privacy concerns} on both \textit{frequency of use} ($\beta = -.23, p < .01$) and \textit{use cases} ($\beta = -.26, p < .01$). Additionally, \textit{ChatGPT proficiency} predicted both \textit{frequency of use} ($\beta = .20, p < .05$) and marginally \textit{use cases} ($\beta = .15, p < .10$). In the group \textbf{with} \textit{organizational policies}, however, none of these effects were observed. These results suggest that individuals in organizations without specific guidance on ChatGPT usage are heavily driven by their personal privacy concerns and rely on their understanding of privacy risks to navigate their use of the tool. In these settings, the absence of formal organizational policies means that employees lack structured directives or training on how to handle ChatGPT responsibly. Consequently, these individuals use their own judgments and concerns about privacy as a primary mechanism for determining their behavior. The two path models can be found in the Appendix (Figs.~\ref{fig:subfig1} and~\ref{fig:subfig2}). 

\paragraph{RQ4.} Lastly, we took a closer look at the group of employees in organizations with \textit{organizational policies} ($n=103$) to determine whether different types of policies organizations employed affected individuals' ChatGPT usage behavior differently. As described before, we differentiated between \textit{usage constraints} (items 3--6) and \textit{user constraints} (items 1 and 2). \textit{Other restrictions} (item 7) were not considered in this analysis, as the cumulative nature of this item makes it hard to interpret. The analysis showed a marginally significant positive effect on the number of \textit{use cases} if the particular type of policies employed by an organization were \textit{usage constraints}. Other than that, no noteworthy differences in the effects of the two types of \textit{organizational policies} were observed. 

\begin{figure*}[t]
\centering
\resizebox{0.8\linewidth}{!}{%
\begin{tikzpicture}
\tikzset{every node/.style={font=\small}}


\draw [->, >=Stealth] (3.0,10.75) -- (5.0,13.75) node[midway, above, fill=white, text centered, outer sep=0pt] {\textbf{.02}} node[midway, below, fill=white, text centered, outer sep=0pt] {\textit{-.07}};

\draw [->, >=Stealth] (3.0,10.75) -- (5.0,7.75) node[midway, above, fill=white, text centered, outer sep=0pt] {\textbf{.07}} node[midway, below, fill=white, text centered, outer sep=0pt] {\textit{-.06}};

\draw [->, >=Stealth] (8.0,13.75) -- (11.0,11.5) node[midway, above, fill=white, text centered, outer sep=0pt] {\textbf{-.02}} node[midway, below, fill=white, text centered, outer sep=0pt] {\textit{-.01}};

\draw [->, >=Stealth] (8.0,13.75) -- (11.0,9.5) node[midway, above, fill=white, text centered, outer sep=0pt] {\textbf{-.10}} node[midway, below, fill=white, text centered, outer sep=0pt] {\textit{-.08}};

\draw [->, >=Stealth] (8.0,7.75) -- (11.0,11.5) node[midway, above, fill=white, text centered, outer sep=0pt] {\textbf{-.06}} node[midway, below, fill=white, text centered, outer sep=0pt] {\textit{-.05}};

\draw [->, >=Stealth] (8.0,7.75) -- (11.0,9.5) node[midway, above, fill=white, text centered, outer sep=0pt] {\textbf{.04}} node[midway, below, fill=white, text centered, outer sep=0pt] {\textit{.05}};

\draw [->, >=Stealth] (1.5,11.5) |- (12.5,15.25) -| (12.5,12.25);

\draw [->, >=Stealth] (1.5,9.75) |- (12.5,6.25) -| (12.5,8.75);

\draw (1.5,10.75) ellipse (1.5cm and 0.75cm);
\node[text centered] at (1.5,10.75) {\parbox[c]{2cm}{\centering \textbf{Usage constr.}\\ \textit{User constr.}}};

\draw (6.5,13.75) ellipse (1.5cm and 0.75cm);
\node[text centered] at (6.5,13.75) {\parbox[c]{2cm}{\centering Integrated\\privacy concerns}};
\node[fill=white, inner sep=4pt, outer sep = 0pt] at (7,15.4) {\textbf{.11}};
\node[fill=white, inner sep=4pt, outer sep = 0pt] at (7,15.0) {\textit{.03}};

\draw (6.5,7.75) ellipse (1.5cm and 0.75cm);
\node[text centered] at (6.5,7.75) {\parbox[c]{2cm}{\centering ChatGPT\\proficiency}};
\node[fill=white,text centered, outer sep = 0pt] at (7,6.5) {\textbf{$.17^{\dagger}$}};
\node[fill=white,text centered, inner sep=5pt, outer sep = 0pt] at (7,6.0) {\textit{.06}};

\draw (12.5, 11.5) ellipse (1.5cm and 0.75cm);
\node[text centered] at (12.5, 11.5) {\parbox[c]{2cm}{\centering Frequency of\\use}};

\draw (12.5,9.5) ellipse (1.5cm and 0.75cm);
\node[text centered] at (12.5,9.5) {\parbox[c]{2cm}{\centering Use cases}};
\end{tikzpicture}%
}%

\caption{Path model for RQ4. Values for \textit{usage constraints} are shown above each path; values for \textit{user constraints} are shown below. Effects of background variables are omitted for readability. $\dagger p<.10$. Model fit (\textit{usage constraints}/\textit{user constraints}): $\chi^2/df=1.00/1.27$, CFI$=1.00/.94$, NNFI$=1.00/.71$, RMSEA$=.00/.05$, and SRMR$=.04/.04$. \authorssource}
\label{fig:CISEM4}
\end{figure*}
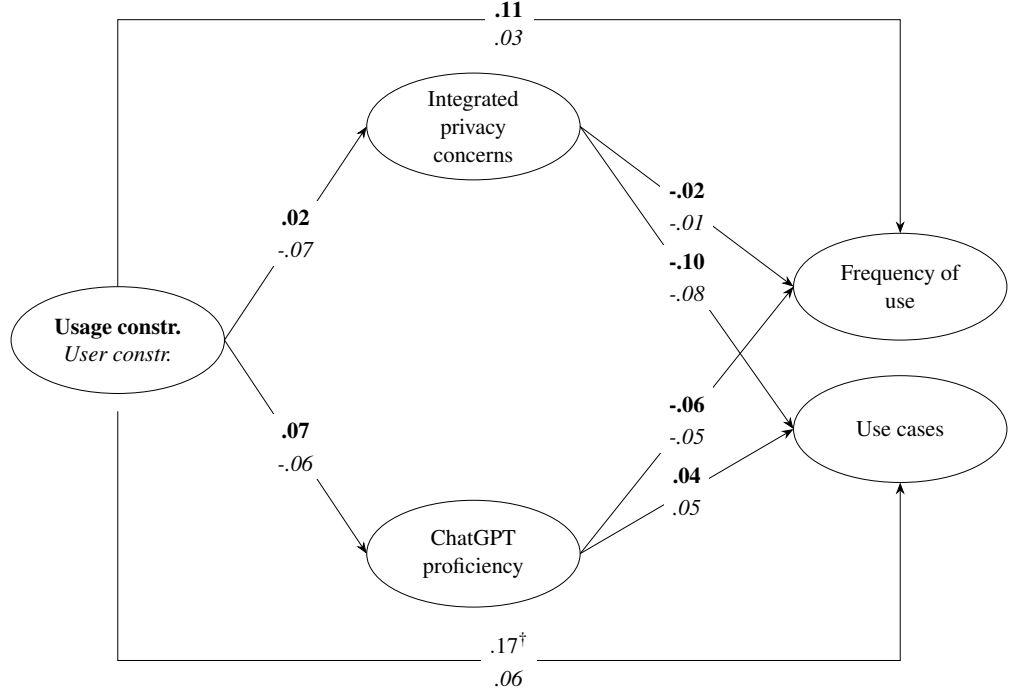

\section{Discussion}
\label{Discussion}

The aim of this study was to investigate how individuals who use ChatGPT in their work perceive potential privacy risks and how these perceptions, along with organizational policies and users' proficiency, shape their ChatGPT usage behavior. In the following, we discuss the findings in relation to our research questions.  

\paragraph{RQ1. Organizational policies and proficiency}  

Regarding \textit{ChatGPT proficiency}, we observed frequent misconceptions, such as the belief that personal information would be automatically anonymized or that OpenAI ensures secure communication channels. Similarly, \cite{Ma25} report that their assessment of mental models for custom GPTs uncovered widespread uncertainty and misconceptions about data flows. For example, some participants believed only OpenAI collected chats to improve its models, others thought GPT creators collected conversations to refine their tools, and a few feared partner companies might receive data for commercial profiling. Participants also reported confusion about login integrations and whether creators could access account credentials. These findings closely mirror our observed misconceptions. While \textit{ChatGPT proficiency} was significantly higher among those participant from organizations with policies in place, however, still with less than 40\% of statements answered correctly on average. These findings show that although organizational policies show a tendency to foster better proficiency, they do not close the existing knowledge gap. This lack of awareness is concerning, as it may lead employees to unintentionally expose sensitive information under the false impression that it is protected. We therefore highlight the need for more targeted education on privacy risks associated with GenAI tools.

\paragraph{RQ2. Policies, privacy concerns, and proficiency on usage behavior}  
Our findings show that organizational policies are not restrictive but instead encourage broader and more versatile use of ChatGPT. Employees in organizations with policies in place applied the tool to a wider range of work-related \textit{use cases}. This finding contradicts the expectation that restrictions could limit creativity or adoption, and instead suggests that policies can provide a framework that empowers safe exploration of new applications. At the same time, \textit{integrated privacy concerns} were negatively related to usage behavior. Participants with higher privacy concerns reported lower \textit{frequency of use} and fewer types of \textit{use cases}. This result is consistent with prior research showing that perceived privacy risks are negatively related to technology adoption \cite{Dinev06_B, Dinev16}, and aligns with findings in the context of chatbots \cite{Eeuwen17, Rese20}. However, it is important to note that some studies have found no significant link between privacy concerns and the intention to adopt or use LLM-based AI technologies~\cite{Vimalkumar21, Menon23}. These conflicting findings suggest that the impact of privacy concerns on technology adoption may vary depending on the specific context or user group, indicating that the relationship between privacy concerns and AI technology usage is not yet fully understood. Interestingly, \textit{ChatGPT proficiency} showed no significant effect on usage behavior, which is surprising given that higher knowledge might be expected to facilitate more active adoption. This points to organizational measures being more decisive than individual knowledge in shaping actual usage behavior.  

\paragraph{RQ3. Moderation by presence or absence of policies}  
The role of \textit{integrated privacy concerns} and \textit{ChatGPT proficiency} in shaping usage behavior differed depending on whether organizational policies were in place. Among participants working in organizations without policies, privacy concerns strongly reduced both frequency of use and range of applications, while proficiency facilitated more active adoption. In contrast, these effects disappeared when policies were present. This suggests that in the absence of organizational guidance, employees rely on their personal judgments about privacy risks, which may limit or drive their adoption depending on individual attitudes and knowledge. When policies are available, they provide structure and guidance that reduce the influence of these individual factors. 

\paragraph{RQ4. Types of organizational policies}  
For RQ4, we examined whether different types of policies have distinct effects on usage behavior. We found little evidence for strong differences. Usage constraints showed a marginal positive effect on the number of use cases, but beyond this, no significant variation emerged between policy types. This suggests that the existence of organizational policies is more important than their specific content. As such, there is little to state about RQ4 in the present study, and future work should explore in more detail how different types of policy interventions may influence adoption patterns.

Taken together, our findings highlight three broader themes. First, despite frequent use of ChatGPT for work tasks, employees generally lack knowledge about how their input data is processed, even though policies improve this to some extent. Second, privacy concerns remain a decisive factor for self restriction, especially in the absence of organizational guidance. Third, organizational policies encourage more versatile usage by providing a framework for safe adoption. In this way, policies help employees navigate privacy risks while also reducing reliance on individual judgments, creating more consistent adoption across different user groups.

\paragraph{Practical Recommendations.} An obvious solution to mitigate privacy risks associated with ChatGPT usage would be to fully block the tool, as organizations like Samsung~\cite{BusinessInsiderSamsungBansEmplyeesFromUsingChatGPT} and JPMorgan~\cite{aibusinessJPMorganJoins} have done in the past. However, this is not an effective strategy. Given the wide availability of alternative GenAI tools like Google’s \textit{Gemini}, employees are likely to circumvent such bans by turning to alternative tools. This introduces even greater risks, as employees might use less vetted tools without proper oversight, exposing the organization to unintended security and privacy vulnerabilities. Another approach might be for organizations to employ local models for GenAI tools, ensuring greater control over data privacy and security. However, this may not be feasible for smaller organizations due to resource constraints. Even with local models, employees might still prefer public tools if they find them more useful or convenient. This presents a significant challenge that organizations must address by implementing well-thought-out policies for GenAI usage.

Organizations must ensure employees engage with GenAI tools mindfully and in a privacy-preserving way. Prior research has shown that employee awareness positively influences adherence to security and privacy policies~\cite{Bulgurcu2010_complianceMIS, Wiant2055_policiesAwareness, Sohrabi2016_complianceAwareness}. Thus, instead of strict restrictions, organizations should focus on policies grounded in employees' awareness of the privacy risks associated with GenAI tools. When employees understand the implications of their actions, they are more likely to comply with organizational guidelines, reducing the risk of privacy breaches. Therefore, establishing educational programs that enhance users' understanding of how GenAI tools handle data is crucial.

To be effective, such educational programs should be mandatory for all employees and designed to go beyond abstract policy communication. They should include practical modules that demonstrate how GenAI tools can be safely and productively integrated into daily work routines. Rather than solely emphasizing risks, the programs should also highlight the opportunities these tools offer for enhancing efficiency, creativity, and decision-making across different roles and departments. For instance, employees could be shown how to use GenAI for drafting routine communications, summarizing reports, generating first drafts of code or documentation, or automating repetitive tasks -- always with a mindful eye on the type of data being shared.

A key focus of the training should be on promoting \textit{mindful use} of GenAI tools. This includes teaching employees to distinguish between low-risk and high-risk use cases, and to assess whether the information they input into GenAI systems might include sensitive, confidential, or personally identifiable data. Where the deployment of a local, on-premises GenAI model is not feasible, due to technical, financial, or organizational constraints, employees should be introduced to privacy-enhancing configurations and tools. For example, browser extensions or client-side filters that strip out sensitive data before sending prompts to cloud-based models that the organization recommends and provides. They should also be made aware of privacy settings offered by GenAI platforms (e.g., disabling chat history, opting out of data logging, or selecting \say{no training} modes where available).

Additionally, organizations could integrate usage guidelines with interactive scenarios that allow employees to test their understanding in realistic workplace situations. By simulating common use cases and demonstrating best practices (e.g., anonymizing input data, avoiding direct sharing of customer records, or cross-checking AI-generated output), such programs can cultivate both digital competence and privacy awareness. This combination of empowerment and precaution supports a culture where employees are encouraged to innovate with GenAI tools -- while upholding the organization’s data protection standards.

However, relying solely on awareness and educational measures may not be sufficient. Organizations also need to implement usage-restricting measures to mitigate potential privacy violations. This approach helps to counteract the effects of the privacy paradox—the discrepancy between individuals' attitudes toward privacy and their actual behaviors. Despite expressing concerns about privacy, people often engage in actions that compromise it~\cite{Dinev06_A,Norberg07} (see Sec.~\ref{sec:related_work}). This paradox can be attributed to factors such as optimistic bias, where individuals believe they are less likely to experience negative outcomes~\cite{Cho10}, or hyperbolic discounting, where immediate benefits are prioritized over potential future risks~\cite{Acquisti03}. Even with well-informed and privacy-aware users, the risks of unintended information disclosure remain. Therefore, it is crucial to implement appropriate policies, tailored to the specific application and scope of the technology, at an early stage.

\section{Limitations and Future Work}
The results of our study must be considered in light of certain limitations, which hold important implications for future research. Acknowledging these limitations, we discuss what may be done in future studies to further validate our findings and which research directions may be suggested based on our study. \par\smallskip

\paragraph{Study scope}
The findings of this study apply only to ChatGPT, as the study focused exclusively on this tool and did not include other GenAI applications. At the same time, the broader concern of users being unaware of how their input data is stored, processed, or potentially reused is not unique to ChatGPT but represents a general challenge for commercial GenAI applications. This issue arises whenever individuals do not have access to a GenAI tool that stores and processes data in a privacy-friendly way, such as locally hosted solutions or paid enterprise versions. Therefore, while the specific findings of this study cannot be generalized beyond ChatGPT, the overarching concern of insufficient awareness of data practices is highly relevant across the wider landscape of commercial GenAI tools. Furthermore, the dynamic nature of AI technology poses challenges for generalizing our findings over time. As AI integrates further into professional and everyday life, the novelty and unfamiliarity that currently characterize adoption will diminish. Future research should reassess these findings as adoption expands across more diverse professional environments with varying organizational characteristics and use cases. 

\paragraph{Sample composition}
Our sample's composition limits generalizability. Since only European participants were included, the findings may not extend to other geographical and cultural contexts. Future studies should address this by incorporating a more diverse sample. Additionally, our sample included few ChatGPT users aged 45 and older, potentially affecting findings for this group. While this demographic distribution aligns with other studies on AI adoption \cite{Vimalkumar21, Eeuwen17}, targeting this age group specifically in future research could provide more nuanced insights. Future studies should aim for a more balanced industry representation to capture broader adoption trends. A further limitation of this study is that the observed effects of organizational policies on both \textit{ChatGPT proficiency} and usage behavior may in part reflect a generally higher level of awareness in organizations that are already more sensitive to privacy risks of GenAI tools, rather than being caused only by the presence of organizational policies. Nonetheless, the analyses show a clear positive impact of policies, even if this aspect should be kept in mind when interpreting the findings. 

\paragraph{Study design}
The study design did not consider whether the participants’ organizations were using ChatGPT Enterprise, where data storage and processing might be handled differently than in the free ChatGPT version, as ChatGPT Enterprise was not widely available at the time of our study. As the adoption of ChatGPT Enterprise increases, it may have significant implications for how organizations approach AI tool integration and should be accounted for in future studies. The \textit{use cases} in our questionnaire were derived from personal experiences and existing literature, but the absence of a comprehensive set of validated use cases may limit the depth of insights into ChatGPT usage. Nevertheless, we aimed to cover a broad range of applications, which we believe effectively capture fundamental aspects of adoption. Regarding the demographic section of our questionnaire, participants reported their occupation, which could have included students reporting “education” as their occupation. While we intended this category to capture professional work within the education sector, this may have been interpreted differently and should be considered when evaluating the applicability of results across different occupation types. Looking ahead to future research, the positive relationship of \textit{organizational policies} with both participants' \textit{ChatGPT proficiency} and their \textit{frequency of use} of ChatGPT and the corresponding variety of \textit{use cases} stands out as particularly noteworthy. Thereby, we did not see any noteworthy differences between different types of policies. This positive role of organizational policies - including both awareness programs and restrictive measures - is not what we initially expected and deserves further investigation in future studies. 

\section{Conclusion}
\label{Conclusion}

ChatGPT and other GenAI tools are about to transform the way we work fundamentally. However, this evolution does not go without certain risks. Our study reveals a significant gap in users' understanding of the privacy risks associated with using ChatGPT despite the continuously growing role ChatGPT plays as a part of their work routines. Accordingly, users' lack of awareness presents substantial risks, especially in the absence of comprehensive organizational guidance. A key finding of our study is that organizational policies which aim at regulating the use of GenAI tools can have a positive impact in this regard. They enhance individuals' knowledge of privacy risks associated with ChatGPT while at the same time encouraging them to frequently use ChatGPT for a variety of purposes in their work. Nevertheless, many organizations do not seem to be aware of the importance of such measures. Only half of our participants reported knowing of any organizational policies governing the use of GenAI tools. To effectively address these challenges, we recommend that organizations develop sensible policies that extend beyond simple restrictions. While regulation is necessary to mitigate risks, the focus should be on training and awareness programs that empower employees by providing them with the knowledge and skills to use GenAI tools in an effective yet responsible way.

\section*{Disclosure of AI Usage}
Language and readability of this paper were enhanced using DeepL Write\footnote{\url{https://www.deepl.com/de/write}}, Grammarly\footnote{\url{https://app.grammarly.com/}}, and ChatGPT\footnote{\url{https://chatgpt.com/}} (version 4o). The authors take full responsibility for the final~content.

\bibliographystyle{apalike}

\bibliography{references}

\clearpage
\onecolumn
\appendix

\section{Correlations}
\label{sec:correlations}

\begin{table}[htbp]
\begin{threeparttable}
\caption{Pearson's correlations for the entire cohort ($N=224$). \authorssource}

\setlength{\tabcolsep}{2pt}      
\renewcommand{\arraystretch}{1.1} 

\begin{tabularx}{\linewidth}{l l *{6}{>{\centering\arraybackslash}X}}
\toprule
& & M & SD & 1 & 2 & 3 & 4 \\
\midrule
1 & Organizational policies &  & &  &  &  &  \\
2 & Integrated privacy concerns & 3.88 & 0.81 & .06 &  &  &  \\
3 & ChatGPT proficiency & 0.36 & 0.18 & .15* & .01 &  &  \\
4 & Frequency of use & 2.50 & 1.00 & -.03 & -.14* & .08 &  \\
5 & Use cases & 2.96 & 1.41 & .23** & -.15* & .13 & .42** \\
\bottomrule
\end{tabularx}

\begin{tablenotes}
\small
\item *$p<.05$, **$p<.01$
\end{tablenotes}

\end{threeparttable}
\end{table}

\begin{table}[htbp]
\begin{threeparttable}
\caption{Pearson's correlations for the group with \textit{organizational policies} regarding GenAI usage in their organizations ($n=103$). \authorssource}
\label{tab:correlation_policies}

\setlength{\tabcolsep}{2pt}      
\renewcommand{\arraystretch}{1.1} 

\begin{tabularx}{\linewidth}{l l *{6}{>{\centering\arraybackslash}X}}
\toprule
& & M & SD & 1 & 2 & 3  \\
\midrule
1 & Integrated privacy concerns & 3.93 & 0.78 &  &  &  \\
2 & ChatGPT proficiency & 0.39 & 0.18 & .01 &  &  \\
3 & Frequency of use & 2.46 & 0.95 &  -.01 & -.06 &  \\
4 & Use cases & 3.31 & 1.48 & -.09 & .05 & .39** \\
\bottomrule
\end{tabularx}
\begin{tablenotes}
\small
\item *$p<.05$, **$p<.01$
\end{tablenotes}
\end{threeparttable}
\end{table}

\begin{table}[htbp]
\begin{threeparttable}
\caption{Pearson's correlations for the group without \textit{organizational policies} regarding GenAI usage in their organizations ($n=121$). \authorssource}
\label{tab:correlation_no_policies}

\setlength{\tabcolsep}{2pt}      
\renewcommand{\arraystretch}{1.1} 

\begin{tabularx}{\linewidth}{l l *{6}{>{\centering\arraybackslash}X}}
\toprule

& & M & SD & 1 & 2 & 3  \\
\midrule
1 & Integrated privacy concerns & 3.83 & 0.83 &  &  &    \\
2 & ChatGPT proficiency & 0.34 & 0.18 & -.01 &  &    \\
3 & Frequency of use & 2.52 & 1.04 & -.23* & .20* &    \\
4 & Use cases & 2.66 & 1.27 &  -.26** & .15 & .49**   \\
\bottomrule
\end{tabularx}
\begin{tablenotes}
\small
\item *$p<.05$, **$p<.01$
\end{tablenotes}
\end{threeparttable}
\end{table}

\section{Path Analyses (RQ3)} 
\label{sec:path_analyses}

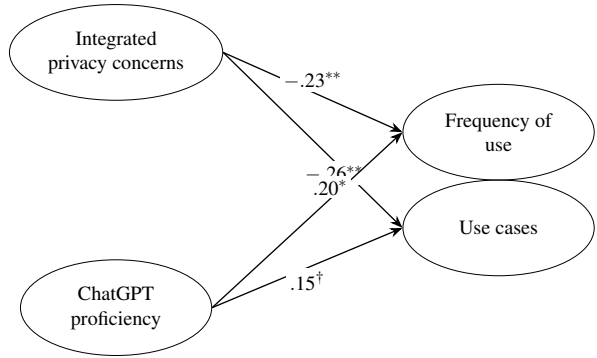
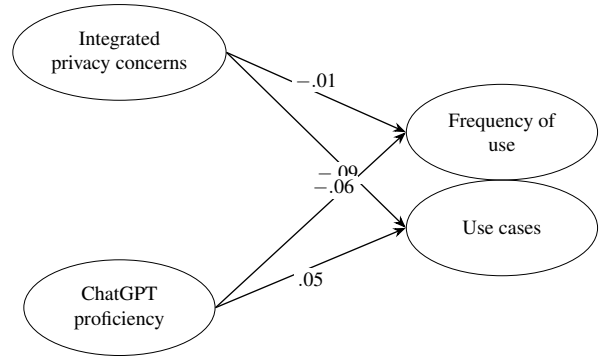
\begin{figure*}[t]
\centering

\begin{subfigure}[t]{0.47\textwidth}
\centering
\resizebox{\linewidth}{!}{%
\begin{tikzpicture}

\node[sem-variable] (concerns) at (0,4)
    {Integrated\\privacy concerns};

\node[sem-variable] (proficiency) at (0,0)
    {ChatGPT\\proficiency};

\node[sem-variable] (frequency) at (6,2.75)
    {Frequency of\\use};

\node[sem-variable] (usecases) at (6,1.25)
    {Use cases};

\draw[sem-path]
    (concerns.east) -- (frequency.west)
    node[sem-coefficient, pos=.50, above] {$-.23^{**}$};

\draw[sem-path]
    (proficiency.east) -- (usecases.west)
    node[sem-coefficient, pos=.50, below] {$.15^{\dagger}$};

\draw[sem-path]
    (concerns.east) -- (usecases.west)
    node[sem-coefficient, pos=.62, below] {$-.26^{**}$};

\draw[sem-path]
    (proficiency.east) -- (frequency.west)
    node[sem-coefficient, pos=.62, above] {$.20^{*}$};

\end{tikzpicture}%
}
\caption{Participants in organizations \textbf{without}
\textit{organizational policies} ($N=121$).}
\label{fig:subfig1}
\end{subfigure}
\hfill
%
\begin{subfigure}[t]{0.47\textwidth}
\centering
\resizebox{\linewidth}{!}{%
\begin{tikzpicture}

\node[sem-variable] (concerns) at (0,4)
    {Integrated\\privacy concerns};

\node[sem-variable] (proficiency) at (0,0)
    {ChatGPT\\proficiency};

\node[sem-variable] (frequency) at (6,2.75)
    {Frequency of\\use};

\node[sem-variable] (usecases) at (6,1.25)
    {Use cases};

\draw[sem-path]
    (concerns.east) -- (frequency.west)
    node[sem-coefficient, pos=.50, above] {$-.01$};

\draw[sem-path]
    (proficiency.east) -- (usecases.west)
    node[sem-coefficient, pos=.50, below] {$.05$};

\draw[sem-path]
    (concerns.east) -- (usecases.west)
    node[sem-coefficient, pos=.62, below] {$-.09$};

\draw[sem-path]
    (proficiency.east) -- (frequency.west)
    node[sem-coefficient, pos=.62, above] {$-.06$};

\end{tikzpicture}%
}
\caption{Participants in organizations \textbf{with}
\textit{organizational policies} ($N=103$).}
\label{fig:subfig2}
\end{subfigure}

\caption{Path models for RQ3 comparing participants in organizations
without organizational policies ($N=121$) and with organizational
policies ($N=103$).
$\dagger p<.10$, ${}^{*}p<.05$, and ${}^{**}p<.01$.
Fit indices are not reported because the models are saturated
($df=0$) and reproduce the observed covariances exactly.
\authorssource}
\label{fig:CISEM23}
\end{figure*}

\small
\section{Questionnaire}
\label{sec:Questionnaire}

\noindent \textit{The following set of questions aims to understand your usage behavior regarding GenAI-based technology, using ChatGPT from OpenAI as an illustrative example.}

\subsection{Organizational Policies}
\textit{To what extent does your employer impose policies on the use of ChatGPT? If limitations exist, kindly specify the scope of these policies or restrictions (please select \textbf{all} that apply to you).}

\begin{itemize}
    \item ChatGPT usage is restricted to specific departments or a circle of users.
    \item Prior to gaining authorization for ChatGPT usage, individuals are required to undergo training or participate in awareness-raising measures.
    \item Usage of ChatGPT is restricted to documents classified up to a specified level of confidentiality (e.g., Public, Internal, Confidential, Strictly Confidential)
    \item Data inputted into ChatGPT must be de-identified or anonymized as a requirement.
    \item Users are reminded that the inclusion of personal information in the input data for ChatGPT is explicitly prohibited.
    \item Limiting the sharing or external use of content generated by ChatGPT.
    \item Other restrictions that have not been mentioned.
    \item No restrictions.
\end{itemize}

\subsection{Integrated Privacy Concerns}
\textit{The following set of questions seeks to assess the significance of privacy to you. For this series of questions, we request your attitude assessment using a five-point Likert-type scale. The scale includes the following rating options: \textbf{(1) Strongly Agree, (2) Agree, (3) Neither Agree nor Disagree (Undecided / Neutral), (4) Disagree or (5) Strongly Disagree.} Kindly share your general perspective on Privacy Concerns and Trust.}

\begin{itemize}
    \item Sharing my personal information does not bother me, so I usually don’t hesitate to provide it. (inverted)
    \item I am concerned that companies collect too much personal information about me.
    \item It bothers me when companies collect personal information in an unclear manner.
    \item I am concerned that companies use my personal information for any other reason of their interest without my knowledge or explicit permission to do so.
    \item It doesn’t bother me when companies sell or share my personal information without prior asking me for my explicit permission to do so. (inverted)
    \item I am not concerned that the information I submit online could be misused in a way I did not foresee. (inverted)
\end{itemize}

\subsection{ChatGPT Proficiency}
\textit{The following set of questions seeks to assess your technical competence in ChatGPT utilization through examining your familiarity with its privacy policy. In this series of questions, we request you to check the accuracy of the subsequently provided statements. We therefore use a three-point response scale including the rating options: \textbf{True (T) / Wrong (W) / Unsure (U)}. Please check the accuracy of the subsequent statements.}

\begin{itemize}
    \item OpenAI automatically captures your IP-address, connections, timestamp and country as well as device information, therefore potentially knowing where, when, and with what device you engage with ChatGPT. 
    \item ChatGPT can tell your preferences and desires, constantly tracking your favorite content types and your interaction with them.
    \item OpenAI is responsible for privacy violations concerning third parties.
    \item OpenAI asserts to ensure the security of online communication channels.
    \item When you upload a document in ChatGPT, your personal information included within the document are collected.
    \item Should there be a need for an identity check, ChatGPT stores the information related to the verification request.
    \item ChatGPT incorporates the content and information you provide into its model for training, indirectly making it accessible to other users.
    \item The default setting allows your data to be used for model training, but you have the option to opt out if you choose.
    \item ChatGPTs Privacy Policy guarantees that personal information are disidentified.
    \item ChatGPT can provide your personal information to third parties without further noticing you.
    \item OpenAI shares personal information with third parties, such as hosting service and cloud service providers, granting them access to process and store the data.
\end{itemize}

\subsection{Frequency of Use}
\textit{How frequently do you use ChatGPT in your professional tasks?}
\begin{itemize}
    \item Daily
    \item Several times a week
    \item Once a week
    \item Less than once a week
\end{itemize}

\subsection{Use Cases}
\textit{Please select the purposes or reasons for using ChatGPT from the following list (please select \textbf{all} that apply to you).}

\begin{itemize}
    \item Code-Assistance: ChatGPT assists me in generating code snippets and/or debugging established code for various tasks/projects.
    \item Email-Assistance: In my workflow, ChatGPT is employed for automatically summarizing emails and crafting responses to the initial message.
    \item Support-Assistance: Quickly analyzing customer inquiries and complaints is facilitated using ChatGPT, allowing for prompt and effective responses.
    \item Meeting-Assistance: Effortlessly summarizing meetings and discussions, along with outlining essential aspects to be addressed, is a seamless process.
    \item Brainstorming and Idea-Assistance: When it comes to getting fresh ideas, handling research tasks, and generating ideas for product development, ChatGPT is my go-to.
    \item Content-Polishment: ChatGPT is my go-to for editing and proofreading to achieve polished content.
    \item Marketing and Content-Generation: ChatGPT streamlines social media and tailors marketing content to campaigns.
    \item Data Analysis Support: With ChatGPT's assistance, data analysis tasks are simplified, and querying for insights is seamless.
    \item Decision Support: When faced with complex decisions, I turn to ChatGPT for support in analyzing information and considering various perspectives.
    \item Language Translation and Communication: ChatGPT aids in language translation, facilitating communication with individuals who speak different languages.
\end{itemize}

\vspace{0.5cm}

\subsection{Demographics}
\textit{In the following, demographic information will be collected. The data are intended solely for research purposes related to the subsequent analysis and interpretation of survey results. Please select the age category that aligns with your current stage of life.}\footnote{Due to the limited sample sizes in the age categories of 45--54, 55--64, and 65+, these were combined into the 45+ category to allow for representative findings.}

\begin{itemize}
    \item 18 - 24
    \item 25 - 34
    \item 35 - 44
    \item 45 - 54
    \item 55 - 64
    \item 65+
\end{itemize}

\noindent \textit{Please indicate your gender identity from the options provided.}

\begin{itemize}
    \item Male
    \item Female
    \item Non-binary
    \item Prefer not to disclose
\end{itemize}

\noindent \textit{Please indicate the highest level of formal education you have completed. \footnote{The categories of formal education originally covered in the questionnaire were combined into three overarching levels of education: \textit{foundational education}, (high school diploma or equivalent, completed vocational training), \textit{advanced education} (technician certification, master craftsman certification) and \textit{academic education} (bachelor's degree, master's degree, and doctoral degree).}}

\begin{itemize}
    \item High School Diploma or Equivalent
    \item Completed Vocational Training
    \item Technician Certification
    \item Master Craftsman Certification
    \item Bachelor's Degree
    \item Master's Degree
    \item Doctoral Degree
    \item None of the named
\end{itemize}

\noindent \textit{Please specify the primary sector of your current occupation or professional activity.}

\begin{itemize}
    \item IT/Technology
    \item Engineering/Manufacturing
    \item Education/Research
    \item Finance/Banking
    \item Business/Consulting
    \item Healthcare/Pharmaceuticals
    \item Arts/Entertainment
    \item Other
\end{itemize}

\clearpage 

\end{document}